\documentclass[12pt]{article}
 
\usepackage{mathptmx}
\usepackage{epsfig}

\usepackage{natbib}
\bibpunct{(}{)}{;}{a}{,}{,}

\usepackage{setspace}

\begin{document}           

\vspace*{-1.0cm}

\begin{center}
  {\large {\bf Rapid solidification of Earth's magma ocean limits
      early lunar recession}}
\vspace{0.5cm}

Jun Korenaga \\
Department of Earth and Planetary Sciences, Yale University \\
P.O. Box 208109, New Haven, CT 06520-8109, USA \\
Email: jun.korenaga@yale.edu \\
Tel: (203) 432-7381; Fax: (203) 432-3134 \\

\end{center}

{\bf Abstract.}  The early evolution of the Earth-Moon system
prescribes the tidal environment of the Hadean Earth and holds the key
to the formation mechanism of the Moon and its thermal
evolution. Estimating its early state by backtracking from the
present, however, suffers from substantial uncertainties associated
with ocean tides. Tidal evolution during the solidification of Earth's
magma ocean, on the other hand, has the potential to provide robust
constraints on the Earth-Moon system before the appearance of a water
ocean. Here we show that energy dissipation in a solidifying magma
ocean results in considerably more limited lunar recession than
previously thought, and that the Moon was probably still at the
distance of $\sim$7-9 Earth radii at the end of solidification.  This
limited early recession aggravates the often overlooked difficulty of
modeling tidal dissipation in Earth's first billion years, but it also
offers a new possibility of resolving the lunar inclination problem by
allowing the operation of multiple excitation mechanisms.

\section{Introduction}
  
Since the pioneering work of George Darwin \cite[]{darwin1879,
  darwin1880}, the tidal evolution of the Earth-Moon system has been
studied by a number of scientists \cite[]{macdonald64, kaula64,
  goldreich66, lambeck75, webb82, touma94, bills99, M_abe01, zahnle15,
  ZL_tian20, farhat22}.  Tidal friction caused by lunar tides slows
down Earth's rotation and results in the gradual recession of the
Moon. It follows that Earth was rotating faster in the past and the
lunar semimajor axis was shorter than its present-day value, $\sim$60
Earth radii ($R_E$).  Estimates on Earth's past rotation rate from
tidal deposits are consistent with such orbital evolution
\cite[]{williams00, coughenour09}, with the oldest inference being
53.5$\pm$0.4~$R_E$ at 1.4~Ga \cite[]{meyers18}.  Going deeper in time
with theory is subject to considerable uncertainties. Ocean tides,
which are overwhelmingly more important than earth tides
\cite[]{lambeck75, webb82}, are sensitive to rotation rate, ocean
depth, seafloor topography, and continental configuration
\cite[]{webb82, M_abe01, motoyama20}, and the Hadean and the Archean
are the period when continents may have grown rapidly
\cite[]{harrison09, rosas18, guo20}, and the ocean likely experienced
non-monotonic volume changes \cite[]{korenaga21a, miyazaki22a}. Whereas the
treatment of ocean tides in recent studies on the Earth-Moon evolution
has become sophisticated \cite[]{tyler21, daher21,farhat22}, the
effects of changing ocean volume as well as changing continental mass
are yet to be considered, so the reliability of such recent results is
unclear for the Hadean and the Archean.

The early evolution of the Earth-Moon distance plays a central role in
various mechanisms proposed for the origin of the lunar inclination,
which in turn is critical to the formation mechanism of the Moon
itself \cite[]{goldreich66, touma98, ward00b, pahlevan15, EMA_chen16,
  cuk16}. Furthermore, the lunar distance controls the magnitude of
tidal dissipation within the Moon, thereby affecting its early thermal
evolution, such as the duration of the lunar magma ocean
\cite[]{elkins-tanton11}. It also determines the amplitude of ocean
tides on Earth, which has important ramifications for the origin of
life. Some of major abiogenesis hypotheses require the presence of
exposed landmasses, on which wet-dry cycles can promote the
polymerization of organic compounds \cite[]{campbell19, damer20}, but
the stable operation of wet-dry cycles would be jeopardized if exposed
landmasses are frequently swamped by huge tides \cite[]{russell06}.

Complications associated with ocean tides can be avoided if we track
the early tidal evolution forward in time, starting from the aftermath
of the Moon-forming giant impact. Though the details of the
Moon-forming giant impact are still actively debated \cite[]{canup12,
  cuk12, asphaug14, rufu17}, the Moon is likely to have formed just
outside the Roche limit, at the distance of $\sim$3-5~$R_E$
\cite[]{salmon12, cuk12}. Tidal dissipation is limited in a completely
molten Earth, but it can be substantial in a solidifying magma ocean,
allowing rapid lunar recession. An order of magnitude calculation
based on an isothermal magma ocean suggests that, while a magma ocean
solidifies within one hundred million years after the giant impact,
the lunar distance may expand to $\sim$40~$R_E$ \cite[]{zahnle15}. The
isothermal approximation is, however, unlikely to represent some
important aspects of the physics of a solidifying magma ocean
\cite[]{abe93b, solomatov93c, abe97, solomatov15tog}. To constrain the
early history of the Earth-Moon system, it becomes essential to use a
more realistic model of Earth's magma ocean. In this paper, we will
show that, in contrast to the earlier suggestion by \cite{zahnle15},
lunar recession during magma ocean solidification is likely to have
been severely limited, reaching only $\sim$7-9 Earth radii, which has
important implications for the origin of the lunar inclination.  The
structure of this paper is as follows. In the Methods section, we
first describe how we parameterize the mantle phase diagram and how we
calculate mantle adiabats. Mantle rheology is then described, and the
timescale of Rayleigh-Taylor instability is estimated, with which the
use of adiabats can be justified for tidal dissipation
calculation. With these preparations, we describe our procedure to
calculate tidal dissipation, thermal evolution, and orbital
evolution. The results of our orbital calculation are then presented,
followed by the discussion of their implications. The main focus of
this paper is to lay out how to take into account the physics of magma
ocean solidification into the tidal dissipation of Earth, and as such,
the other aspects of tidal evolution are kept simple.

\section{Methods}

\subsection{Phase diagram}

Based on the experimental and theoretical studies of high-pressure
mantle melting \cite[]{zhang94, fiquet10, miyazaki19a}, the
mantle phase diagram is parameterized using cubic splines as follows:
\begin{eqnarray}
  T_s(z) &=&  \sum_{i=1}^{3} [H(z-z^i)-H(z-z^{i+1})] \,
  S(T_s^i,T_s^{i+1},b_s^i,b_s^{i+1},z^i,z^{i+1},z), \\
  T_l(z) &=& \sum_{i=1}^{3} [H(z-z^i)-H(z-z^{i+1})] \,
  S(T_l^i,T_l^{i+1},b_l^i,b_l^{i+1},z^i,z^{i+1},z),
\end{eqnarray}
where $T_s$ and $T_l$ are solidus and liquidus temperatures,
respectively, in Kelvin, $z$ is depth in km, $H()$ is the Heaviside step
function, and $S()$ is the cubic spline function. The spline function
is defined as
\begin{eqnarray}
  S(T_i,T_{i+1},b_i,b_{i+1},z_i,z_{i+1},z) &=&
  \frac{b_{i+1}(z-z_{i})^3}{6 h_i} + \frac{b_i(z_{i+1}-z)^3}{6 h_i} \nonumber \\
 & & + \left[\frac{T_{i+1}}{h_i}-\frac{b_{i+1} h_i}{6}\right] (z-z_i)
  + \left[\frac{T_{i}}{h_i}-\frac{b_{i} h_i}{6}\right] (z_{i+1}-z),
\end{eqnarray}
where $h_i = z_{i+1}-z_i$, $z_i$ is the $i$th knot, and $T_i$ and
$b_i$ are the values of the function and its second derivative at the
$i$th knot. We use four knots, located at the depths of 0, 410, 670,
and 2900~km, with $T_s^i = \{1273, 2323, 2473, 3985\}$, $b_s^i =
\{-10^{-3}, -7\times10^{-3}, -10^{-3}, 3\times10^{-4}\}$, $T_l^i =
\{1973, 2423, 2723, 5375\}$, and $b_l^i = \{-10^{-3}, -7\times10^{-3},
-2.5 \times 10^{-3}, 5\times10^{-4}\}$. 

We also define $T_{40}$ as the temperature at which the melt fraction
is 40\%, and it is parameterized as
\begin{equation}
  T_{40}(z) = T_s(z) + f_{40}(z) [T_l(z) - T_s(z)],
\end{equation}
where $f_{40}(z)$ is linearly interpolated from 0.4 at 0~km, 0.4 at
410~km, 0.6 at 670~km, and 0.845 at 2900~km. That is, the melt
fraction is more nonlinearly distributed between the solidus and the
liquidus in the lower mantle \cite[]{miyazaki19b}. This is
our standard choice of the mantle phase diagram and is referred to as
`MK19'. For comparison, we also use the version called `const', in
which $f_{40}$ is 0.4 throughout the mantle. At a given depth, the
melt fraction between $T_s$ and $T_{40}$ is assumed to be distributed
linearly from 0 to 0.4, and that between $T_{40}$ and $T_l$ is also
linearly from 0.4 to 1.0.

\subsection{Mantle adiabat \label{sec:adiabat}}

When temperature is above the liquidus or below the solidus, a mantle
adiabat for a given surface temperature is calculated simply by
integrating the isentropic thermal gradient
\begin{equation}
  \left( \frac{\partial T}{\partial P} \right)_S^0
  = \frac{\alpha(T,P) T}{\rho(T,P) C_P(T,P)},
\end{equation}
where $T$ is temperature, $P$ is pressure, $\alpha$ is thermal
expansivity (K$^{-1}$), $\rho$ is density (kg~m$^{-3}$), and $C_p$ is
isobaric specific heat (J~kg$^{-1}$~K$^{-1}$). Based on 
the thermodynamic calculations conducted by \cite{miyazaki19b},
these material properties are parameterized as
\begin{eqnarray}
  \alpha(T,P) &=& 3.622 \times 10^{-5} \exp(-2.377\times 10^{-5} T
  -0.0106 P), \\
  \rho(T,P) &=& 2870 - 0.082 T + 162 P^{0.58}, \\
  C_P(T,P) &=& 627+0.411 T - 0.211 P,
\end{eqnarray}
where $T$ is in Kelvin and $P$ is in GPa.

When temperature is between the solidus and the liquidus, the effect
of the latent heat of fusion must be taken into account as
\begin{equation}
  \left( \frac{\partial T}{\partial P} \right)_S
  = \left( \frac{\partial T}{\partial P} \right)_S^0
  + \frac{H_f}{C_P} \left(\frac{\partial \phi}{\partial P}\right)_S,
\end{equation}
where $H_f$ is the latent heat of fusion and $\phi$ is the melt
fraction.  The term $(\partial \phi/\partial P)_S$ can be obtained by
requiring self-consistency with $\phi(T,P)$ given by the chosen mantle
phase diagram \cite[]{langmuir92}. To approximate the adiabat
calculations of \cite{miyazaki19b}, which are based on Gibbs
energy minimization, the latent heat of fusion is linearly
interpolated from $6 \times 10^5$~J~kg$^{-1}$ at 0~GPa and $9 \times
10^6$~J~kg$^{-1}$ at 136~GPa. 

As seen in Figure~\ref{fig:adiabat}a, upon cooling, the mantle adiabat crosses the
liquidus at the base of the mantle. Crossing the liquidus first at
some mid-mantle depth, i.e., the formation of a basal magma ocean, has
been suggested on the basis of the thermodynamics of simple silicate
melt \cite[]{labrosse07b, mosenfelder09, stixrude09}, but its
possibility appears to be small for mantle-like melt compositions
\cite[]{miyazaki19b}.

\subsection{Rheology}

The rheology of a magma ocean is a strong function of the melt
fraction, and when $\phi > \phi_c$, where $\phi_c$ is the critical
melt fraction, we simply assume that viscosity takes the value of melt
viscosity, $\eta_L$. The viscosity of a solid-melt mixture increases
by a few orders of magnitude when $\phi$ approaches $\phi_c$ from the
above \cite[]{abe93b, solomatov15tog}, but such an increase is
considerably smaller than a viscosity jump associated with rheological
transition at $\phi = \phi_c$. Our simplified treatment does not
affect our calculation of tidal dissipation, in which a layer with
$\phi>\phi_c$ is treated as a fluid. We set $\phi_c$ to the
traditional value of 0.4 \cite[]{abe93b, solomatov15tog} for our main
results.  An experimental study with olivine aggregates suggests
$\phi_c \sim 0.3$ \cite[]{scott06}, and experiments with other
silicate systems suggest that $\phi_c$ may vary from 0.2 to 0.6
\cite[]{costa09}. As shown later, however, the effect of varying
$\phi_c$ on lunar recession is minor. Melt viscosity is 0.1~Pa~s for
peridotitic melt composition \cite[]{dingwell04}; this is most
appropriate for Earth's magma ocean, so it is used in the reference
case.  At the late stage of magma ocean solidification, higher melt
viscosity corresponding to basaltic melt composition (5-10~Pa~s
\cite[]{mcbirney93}) may become more appropriate.

When $\phi<=\phi_c$, the viscosity of a solid-melt
mixture (or just a solid when $\phi=0$) is calculated as
\begin{equation}
  \eta(z,\phi,T) = \eta_S(z) \exp(-\alpha_{\eta} \phi)
  \exp\left(\frac{E}{RT}-\frac{E}{RT_s(z)} \right),
\end{equation}
where $\eta_s(z)$ is the background solid viscosity at the solidus,
$E$ is the activation energy, and $R$ is the universal gas
constant. Experimental studies suggest $\alpha_{\eta} \sim 26$ for
diffusion creep and $\alpha_{\eta} \sim 31-32$ for dislocation creep
\cite[]{mei02, scott06}. For the reference case, $\alpha_{\eta}$ is set
to 26. The activation energy is set to 300~kJ~mol$^{-1}$
\cite[]{korenaga06}. For the reference case, $\log_{10} \eta_s$ is
linearly interpolated from 19 at the top and 20 at the bottom of the
mantle; this background viscosity is based on an extrapolation from
the present-day viscosity structure \cite[]{miyazaki19b}. For softer
background viscosity, we interpolate $\log_{10} \eta_s$ from 18 and
19, and for harder background viscosity, from 20 to 21. The viscosity
of the present-day convecting mantle is lowered by its water content
\cite[]{hirth96}, and the solid phase coexisting with the melt phase is
dry because water is mostly partitioned into the melt phase. This
consideration is absent in the extrapolation done by
\cite{miyazaki19b}, so the harder background viscosity may be
more appropriate \cite[]{karato86b}.

For viscoelastic rheology used in the calculation of tidal
dissipation, we use the temperature-dependent model of
\cite{takei17}, which offers a more complete treatment of
viscoelasticity compared to earlier attempts \cite[]{faul05,
  I_jackson10}. In this model, the complex compliance is parameterized
as
\begin{equation}
  \tilde{J}(\omega) = J_1(\omega) + i J_2(\omega),
\end{equation}
with
\begin{equation}
  J_1 = J_U \left\{1 + \frac{A_B(p^{\prime})^{\alpha_B}}{\alpha_B}
    + \frac{\sqrt{2\pi}}{2} A_P \sigma_P \left[
      1 - \mbox{erf} \left( \frac{\ln (\tau^{\prime}_P/p^{\prime})}{\sqrt{2} \sigma_P}
      \right)\right]\right\}
\end{equation}
and
\begin{equation}
  J_2 = J_U \left\{p^{\prime} + \frac{\pi}{2}\left[
      A_B(p^{\prime})^{\alpha_B}
      + A_P \exp\left(-\frac{[\ln(\tau^{\prime}_P/p^{\prime})]^2}{2\sigma_P^2}
      \right)\right]\right\},
\end{equation}
where $\omega$ is the tide-raising frequency, $J_U$ is the unrelaxed
compliance ($=1/\mu_E$, where $\mu_E$ is the elastic shear modulus),
and $p^{\prime}$ is the period normalized by the Maxwell relaxation
time ($\tau_M = \eta/\mu_E$), i.e., $p^{\prime} = (\omega
\tau_M)^{-1}$, and other parameters are defined as in
\cite{yamauchi16} and \cite{takei17}.  The elastic shear modulus is
calculated as the Voigt-Reuss-Hill average of the shear moduli of the
solid and melt phases.  The shear modulus of the solid phase is
calculated as a function of depth from the Preliminary Reference Earth
Model (PREM \cite[]{dziewonski81}), and that of the melt phase (at the
infinite frequency limit) is assigned a value of 10~GPa
\cite[]{webb95}.  The bulk viscosity is practically zero
\cite[]{schubert01book}, so we assume that the bulk modulus remains
purely elastic. The complex shear modulus, $\tilde{\mu}$, is simply
the reciprocal of the complex compliance.

The temperature-dependent model of \cite{takei17} may be
considered as a modified Andrade model, characterized by a broad
relaxation spectrum. For comparison, we also consider a Maxwell body,
for which the complex shear modulus is calculated as
\cite[]{tobie05}:
\begin{equation}
  \tilde{\mu}(\omega) = \frac{\mu_E \omega^2 \eta^2}{\mu_E^2 +
    \omega^2\eta^2} + i \frac{\mu_E^2 \omega \eta}{\mu_E^2 +
    \omega^2\eta^2}.
\end{equation}
Tidal dissipation, and thus lunar recession, is more limited with
Maxwell body.

\subsection{Rayleigh-Taylor instability}

When a magma ocean solidifies, it solidifies upwards from the bottom
\cite[]{solomatov15tog}, and because viscosity increases by many orders
of magnitude at the rheological transition at $T=T_{40}$, the thermal
structure of a solidifying magma ocean initially follows $T_{40}(z)$
(e.g., Figure~2a of \cite{miyazaki19b}). However, $dT_{40}/dP$
is greater than the adiabatic gradient, meaning that the solid-melt
mixture lying on $T_{40}(z)$ is gravitationally unstable, susceptible to the
Rayleigh-Taylor instability \cite[]{solomatov15tog, miyazaki19b}. The
time scale of this instability may be estimated as \cite[]{turcotte02}
\begin{equation}
  \tau_{\mbox{\scriptsize{RT}}} = \frac{26 \eta_{\mbox{\scriptsize eff}}}{\Delta \rho g L},
\end{equation}
where $\eta_{\mbox{\scriptsize eff}}$ is the effective viscosity,
$\Delta \rho$ is the density difference across the superadiabatic
layer, $g$ is gravitational acceleration, $L$ is the thickness of the
superadiabatic layer. After the Rayleigh-Taylor instability, the
thermal structure becomes adiabatic, so the effective viscosity
relevant to this transition may be calculated from the logarithmic
average of two viscosity profiles, one along $T_{40}(z)$ and the other
along the final adiabat. The density difference can be calculated from
the difference between these two thermal profiles, using density and
thermal expansivity parameterized in the subsection~``Mantle adiabat''.  Some
representative results are shown in Figure~\ref{fig:RT}. For the
reference case, the time scale becomes less than one year as soon as
the surface temperature drops to $\sim$2800~K and remains below
100~years until the end of the magma ocean phase. Even with the harder
background viscosity, the time scale is less than $10^3$ years, which
is still two orders of magnitude shorter than the time scale of magma
ocean solidification, justifying the assumption of an adiabatic
thermal structure throughout solidification.

Using adiabats calculated with a chosen mantle phase diagram is
equivalent to assuming equilibrium crystallization, i.e., the
composition of a solid-melt mixture does not evolve during magma ocean
solidification. Fractional crystallization is equally likely
\cite[]{solomatov15tog}, and it has been suggested that fractional
crystallization is necessary for the rapid sequestration of
atmospheric carbon in the early Earth \cite[]{miyazaki22a}. The
assumption of equilibrium crystallization in this study is merely for
the sake of simplicity, as the mode of crystallization has only
limited influence on the time scale of magma ocean solidification
\cite[]{miyazaki19b}. We also do not consider the temporal evolution of
melt fraction caused by melt percolation. Melt percolation in the
solid phase reduces the melt fraction in one part, which increases
viscosity thus reduces tidal dissipation, and creates a melt layer in
another part, which does not contribute to tidal dissipation. Thus,
the assumption of no melt percolation in a solidifying magma ocean
maximizes its tidal dissipation.

\subsection{Tidal dissipation}

The tidal deformation of a compressible self-gravitating Earth caused
by the Moon is calculated by solving the following coupled ordinary
differential equations \cite[]{takeuchi72, sabadini16book}:
\begin{equation}
  \frac{dy_i(r,\omega)}{dr} =
  \sum_{j=1}^6 A_{ij} (r,\omega) y_j(r,\omega),
\end{equation}
where $r$ is the radius, $\omega$ is the tide-raising frequency, $y_1$
and $y_2$ are the radial and tangential displacements, respectively,
$y_3$ and $y_4$ are the radial and tangential stresses, respectively,
$y_5$ is the gravitational potential, and $y_6$ is the potential
stress. Our formulation is identical to that of
\cite{sabadini16book} except that we do not neglect inertia,
which is important when considering a fluid layer. The non-zero
elements of the matrix $A$ are:
\begin{eqnarray}
  A_{11} &=& -\frac{2 \tilde{\lambda}}{r \beta},  \\
  A_{12} &=& \frac{l(l+1) \tilde{\lambda}}{r\tilde{\beta}}, \\
  A_{13} &=& \frac{1}{\tilde{\beta}}, \\
  A_{21} &=& -\frac{1}{r}, \\
  A_{22} &=& \frac{1}{r}, \\
  A_{24} &=& \frac{1}{\tilde{\mu}}, \\
  A_{31} &=& \frac{4}{r}\left(\frac{3K\tilde{\mu}}{r\tilde{\beta}}
  -\rho_0 g \right) - \rho_0 \omega^2, \\
  A_{32} &=& \frac{l(l+1)}{r}\left(\rho_0 g -
  \frac{6K\tilde{\mu}}{r\tilde{\beta}} \right), \\
  A_{33} &=& -\frac{4\tilde{\mu}}{r\tilde{\beta}}, \\
  A_{34} &=&  \frac{l(l+1)}{r}, \\
  A_{35} &=&  -\frac{\rho_0 (l+1)}{r}, \\
  A_{36} &=& \rho_0, \\
  A_{41} &=& \frac{1}{r}\left(\rho_0 g -
  \frac{6K\tilde{\mu}}{r\tilde{\beta}} \right), \\
  A_{42} &=& \frac{2\tilde{\mu}}{r^2} \left[ l(l+1)
    \left( 1 + \frac{\tilde{\lambda}}{\tilde{\beta}}\right)-1 \right]
  - \rho_0 \omega^2, \\
  A_{43} &=& -\frac{\tilde{\lambda}}{r \tilde{\beta}},  \\
  A_{44} &=& -\frac{3}{r}, \\
  A_{45} &=& \frac{\rho_0}{r}, \\
  A_{51} &=& -4\pi G \rho_0, \\
  A_{55} &=& -\frac{l+1}{r},  \\
  A_{56} &=& 1, \\
  A_{61} &=& -\frac{4\pi G \rho_0 (l+1)}{r}, \\
  A_{62} &=& \frac{4\pi G \rho_0 (l+1)}{r}, \\
  A_{66} &=& \frac{l-1}{r},
\end{eqnarray}
where $l$ is the order of spherical harmonics, $G$ is the universal
gravitational constant, $\rho_o$ is the reference density, $g$ is the
corresponding gravitational acceleration, $K$ is the bulk modulus, and
$\tilde{\lambda}$ and $\tilde{\beta}$ are defined as
\begin{eqnarray}
  \tilde{\lambda} &=& K - \frac{2}{3} \tilde{\mu}, \\
  \tilde{\beta} &=& \tilde{\lambda} + 2 \tilde{\mu}.
\end{eqnarray}
We use PREM to calculate $\rho_0$, $g$, and $K$. 

A region with $\phi>\phi_c$ is regarded as a fluid layer, in which the
elastic shear modulus vanishes and the tangential stress $y_4$ is
uniformly zero. Because $A_{24}$ diverges, $y_2$ is obtained by making
use of the equation for $dy_4/dr$ \cite[]{takeuchi72}.

In case of the semidiurnal lunar tide ($l=2$), these differential
equations are subject to the following boundary conditions at the
surface: $y_3=y_4=0$ and
\begin{equation}
  y_6 = -\frac{(2l+1)g_s}{4} \frac{M_M}{M_E}
  \left( \frac{R_E}{a} \right)^3,
\end{equation}
where $g_s$ is the surface gravity, $M_M$ and $M_E$ are the masses
of the Moon and Earth, respectively, $R_E$ is the Earth radius, and
$a$ is the lunar distance, and at the core-mantle
boundary (CMB; $r=r_C$) \cite[]{sabadini16book}:
\begin{equation}
  \left( \begin{array}{c} y_1 \\ y_2 \\ y_3 \\ y_4 \\ y_5 \\ y_6 \end{array} \right)
  = \left(  \begin{array}{ccc}
    -\psi_C/g_C & 0 & 1 \\
    0 & 1 & 0 \\
    0 & 0 & g_C/\rho_C \\
    0 & 0 & 0 \\
    \psi_C & 0 & 0 \\
    q_C & 0 & 4\pi G \rho_C \end{array} \right)
    \left( \begin{array}{c} C_1 \\ C_2 \\ C_3 \end{array} \right),
\end{equation}
where $g_C$ is the gravity at CMB, $\rho_C$ is the core density (we
assume a uniform density for the core), 
\begin{eqnarray}
  \psi_C &=& r_C^l, \\
  q_C &=& (2l+1) r_C^{l-1} - 4\pi G \rho_C \frac{r_c^l}{g_C},
\end{eqnarray}
and $C_i$ are the constants to be determined to satisfy the boundary
conditions at the surface. We use the shooting method with the
4th-order Runge-Kutta integration scheme to solve the above
differential equations \cite[]{tobie05}.

To calculate tidal dissipation, the strain rate tensor is first
obtained from $y_1$ and $y_2$ as
\begin{eqnarray}
  \tilde{\epsilon}_{11} &=& \frac{\partial y_1}{\partial r} Y_{lm}, \\
  \tilde{\epsilon}_{22} &=& \frac{y_1}{r} Y_{lm}
  + \frac{y_2}{r} \frac{\partial^2 Y_{lm}}{\partial^2 \theta}, \\
  \tilde{\epsilon}_{33} &=& \frac{y_1}{r} Y_{lm}
  + \frac{y_2}{r} \left( \cot\theta \frac{\partial Y_{lm}}{\partial \theta}
  + \frac{1}{\sin^2\theta} \frac{\partial^2 Y_{lm}}{\partial^2 \phi} \right), \\
  \tilde{\epsilon}_{23} &=& \frac{y_2}{r \sin\theta} \left(
  \frac{\partial^2 Y_{lm}}{\partial \theta \partial \phi}
  - \cot\theta \frac{\partial Y_{lm}}{\partial \phi}\right), \\
  \tilde{\epsilon}_{13} &=& \frac{1}{2 \sin\theta} \left(
  \frac{y_1}{r} + \frac{\partial y_2}{\partial r} - \frac{y_2}{r} \right)
  \frac{\partial Y_{lm}}{\partial \phi}, \\
  \tilde{\epsilon}_{12} &=& \frac{1}{2} \left(
  \frac{y_1}{r} + \frac{\partial y_2}{\partial r} - \frac{y_2}{r} \right)
  \frac{\partial Y_{lm}}{\partial \theta},
\end{eqnarray}
where $\theta$ is the colatitude, $\phi$ is the longitude, and
$Y_{lm}$ is the unnormalized spherical harmonics of degree $l$ and
order $m$. In case of a semidiurnal tide, $l=m=2$. Note that the
symbol $\phi$ is also used for melt fraction in this paper, but
which one is used should be clear from the context. Then, the stress
tensor is calculated as
\begin{equation}
  \tilde{\sigma}_{ij} = 2 \tilde{\mu} \tilde{\epsilon}_{ij}
  + \left[ K - \frac{2}{3} \tilde{\mu}\right] \tilde{\epsilon}_{ij} \delta_{ij}.
\end{equation}
Finally, tidal dissipation is obtained by evaluating the following volume
integral \cite[]{tobie05}:
\begin{equation}
  \dot{E}_{\mbox{\scriptsize earth}} = \frac{\omega}{2}
  \int_{r_C}^{R_E} \left( \int_0^{2\pi} \int_0^{\pi} \left[
    \mbox{Im}(\tilde{\sigma}_{ij}) \mbox{Re}(\tilde{\epsilon}_{ij})
    - \mbox{Re}(\tilde{\sigma}_{ij}) \mbox{Im}(\tilde{\epsilon}_{ij}) \right]
  d\theta d\phi \right)dr. \label{eq:dissipation}
\end{equation}
We validate our code using the case of ``liquid Fe-FeS core + silicate
mantle'' in \cite{tobie05}.

The dissipation of earth tide, as a function of the lunar distance, is
shown in Figure~\ref{fig:adiabat}c for the `MK19' phase diagram and in
Figure~\ref{fig:adiabat2}c for the `const' phase diagram. As $T_{40}$
is lower for the latter, significant tidal dissipation takes place
only after the potential temperature becomes lower than $\sim$2550~K. 

Following \cite{zahnle15}, tidal $Q$ is calculated
as $Q_{\mbox{\scriptsize tidal}} = \dot{E}_{\mbox{\scriptsize
    pot}}/\dot{E}_{\mbox{\scriptsize earth}}$, with \cite[]{kaula68book} 
\begin{equation}
  \dot{E}_{\mbox{\scriptsize pot}} = \frac{3}{4} \frac{\omega k_2 G M_M^2 R_E^5}{a^6},
\end{equation}
where $k_2$ is the Love number for Earth ($=0.3$). During the magma
ocean phase, there is no surface water ocean, so there is no need to
consider the ocean tide. At the early phase of magma ocean
solidification when the adiabat is still above $T_{40}$,
$\dot{E}_{\mbox{\scriptsize earth}}$ is calculated as
$\dot{E}_{\mbox{\scriptsize pot}}/Q_{\mbox{\scriptsize
    tidal}}^{\mbox{\scriptsize max}}$ where $Q_{\mbox{\scriptsize
    tidal}}^{\mbox{\scriptsize max}}$ is set to $10^4$ in the
reference case.Tidal dissipation in an entirely fluid sphere is
estimated to have $Q_{\mbox{\scriptsize tidal}}$ of $10^4$ to $10^6$
\cite[]{zahnle15}. Even after part of the magma ocean starts to
experience the rheological transition, dissipation due to earth tide
during the magma ocean phase is set to the the value given by this
$Q_{\mbox{\scriptsize tidal}}^{\mbox{\scriptsize max}}$ if the
dissipation based on equation~(\ref{eq:dissipation}) is smaller.

\subsection{Thermal evolution}

The thermal evolution of Earth is tracked by solving the following
energy balance:
\begin{equation}
  C_m(T_p) \frac{dT_p}{dt} = \dot{E}_{\mbox{\scriptsize earth}}(t)
  + H_m(t) + Q_c(t) - Q_s(t),
\end{equation}
where $t$ is time, $T_p$ is the potential temperature of the mantle,
$C_m(T_p)$ is the heat capacity of the mantle as a function of its
potential temperature, $H_m$ is radiogenic heating within the mantle,
$Q_c$ is the core heat flux, and $Q_s$ is the surface heat
flux. Time-dependent radiogenic heating is scaled to reproduce the
present-day Urey ratio of 0.3 \cite[]{korenaga06}, and the core heat
flux is linearly interpolated from 15~TW at the beginning and 10~TW at
the present \cite[]{orourke17}. Being much smaller than other terms,
however, the effects of radiogenic heating and core heat flux are
negligible during the magma ocean phase.  To take into account the
release of the latent heat of crystallization upon cooling, the heat
capacity function is constructed from the material properties used to
calculate mantle adiabats (subsection~``Mantle adiabat'',
Figure~\ref{fig:Cm}). In the petrology literature, the mantle
potential temperature is defined as the temperature of the mantle
adiabatically brought up to the surface without melting
\cite[]{mckenzie88}, but in the magma ocean literature, the effect
of melting is not suppressed \cite[]{solomatov15tog} as doing so is
obviously awkward. We follow the latter convention in this study.  In
all cases, we start with the initial potential temperature of 4500~K;
model results are not very sensitive to the initial temperature as
long as it is above the liquidus.

Surface heat flux from a magma ocean is calculated as \cite[]{solomatov15tog}:
\begin{equation}
  Q_s = 0.089 A_E k_m \left( \frac{\alpha_s \rho_s^2 C_p g_s}{k_m \eta_L}
  \right)^{1/3} (T_p-T_s)^{4/3}, \label{eq:soft}
\end{equation}
where $A_E$ is the surface area of Earth, $k_m$ is the melt thermal
conductivity (2~W~K$^{-1}$~m$^{-1}$), $\alpha_s$ is the melt thermal
expansivity ($5 \times 10^{-5}$~K$^{-1}$), $\rho_s$ is the surface
melt density (3000~kg~m$^{-3}$), $C_p$ is the specific heat
($10^3$~J~kg$^{-1}$~K$^{-1}$), $g_s$ is the surface gravity, and $T_s$
is the surface temperature. Earth's magma ocean is considered to be
covered by a dense ($\sim$100~bar) CO$_2$-rich atmosphere
\cite[]{zahnle07}, and based on the 1-D radiative-convective atmosphere
modeling of \cite{miyazaki22a}, the surface temperature is
parameterized as
\begin{equation}
  T_s = c_0 + c_1 T_p,
\end{equation}
where $c_0 = 546$ and $c_1 = 0.63$. Equation~(\ref{eq:soft}) is for
soft turbulence, and heat flow scaling for hard turbulence, which may
be more appropriate for magma oceans, predicts even higher heat flow
\cite[]{solomatov15tog}. Here we choose the scaling for soft turbulence
to provide a conservative estimate on the time scale of magma ocean
solidification. The magma ocean stage ends when the mantle potential
temperature reaches $T_{40}(z=0)$ ($\sim$1613~K), and the surface
temperature is assumed to collapse to 500~K. Before subsolidus mantle
convection eventually takes over, residual melt in the shallow upper
mantle migrates upward and degas, forming a water ocean
\cite[]{miyazaki22a}.

Owing to contrasting solubilities in magma \cite[]{elkins-tanton08},
most of carbon dioxide is in the atmosphere whereas most of water is
in the magma ocean, and this situation is maintained during magma
ocean solidification \cite[]{hier-majumder17, miyazaki22a}. Given the
likely water budget of the bulk Earth ($\sim$1-3 ocean)
\cite[]{hirschmann09, korenaga17b}, therefore, the amount of water in
the atmosphere is too small to form a water ocean during the magma
ocean phase, and the Komabayashi-Ingersoll limit (or the runaway
greenhouse threshold) is not relevant because the limit requires the
coexistence of two phases (water vapor and liquid water)
\cite[]{pierrehumbert11book}.

\subsection{Orbital evolution}

To track the lunar distance, we adopt the simplest assumption that the
Moon evolves away from Earth maintaining a circular orbit
\cite[]{zahnle15}.  We also assume zero obliquity and zero
inclination, so we are concerned only with the semidiurnal tide with
the angular frequency of $2(\Omega-n_1)$, where $\Omega$ is the
angular velocity of Earth's rotation and $n_1$ is the angular velocity
of the Moon on its orbit. This facilitates the comparison of our
results with those of \cite{zahnle15}, who made the same
assumptions. We have tested with the effect of non-zero lunar
inclination on tidal dissipation, with negligible differences.

With these assumptions, the orbital evolution is described by
\begin{equation}
  \frac{d\Omega}{dt} = - \frac{\displaystyle \dot{E}_{\mbox{\scriptsize{tide}}}}
  {\displaystyle I_E \Omega + \frac{G M_E M_M I_E}{a (L-I_E\Omega)}},
\end{equation}
and
\begin{equation}
  \frac{da}{dt} = - \frac{2 I_E a}{L-I_E \Omega} \frac{d\Omega}{dt},
\end{equation}
where $I_E$ is Earth's moment of inertia and $L$ is the total angular momentum
of the Earth-Moon system:
\begin{equation}
  L = I_E \Omega + M_M \sqrt{G(M_E+M_M)a}.
\end{equation}
The lunar orbital angular velocity is updated as
\begin{equation}
  n_1 = \sqrt{G(M_E+M_M)/a^3}.
\end{equation}

Time-stepping is controlled so that a relative change in the lunar
distance change is less than 0.1~\% and that in mantle potential
temperature is less than 2~K at every time step.

\section{Results}

\begin{figure}
\begin{center}
\epsfig{file=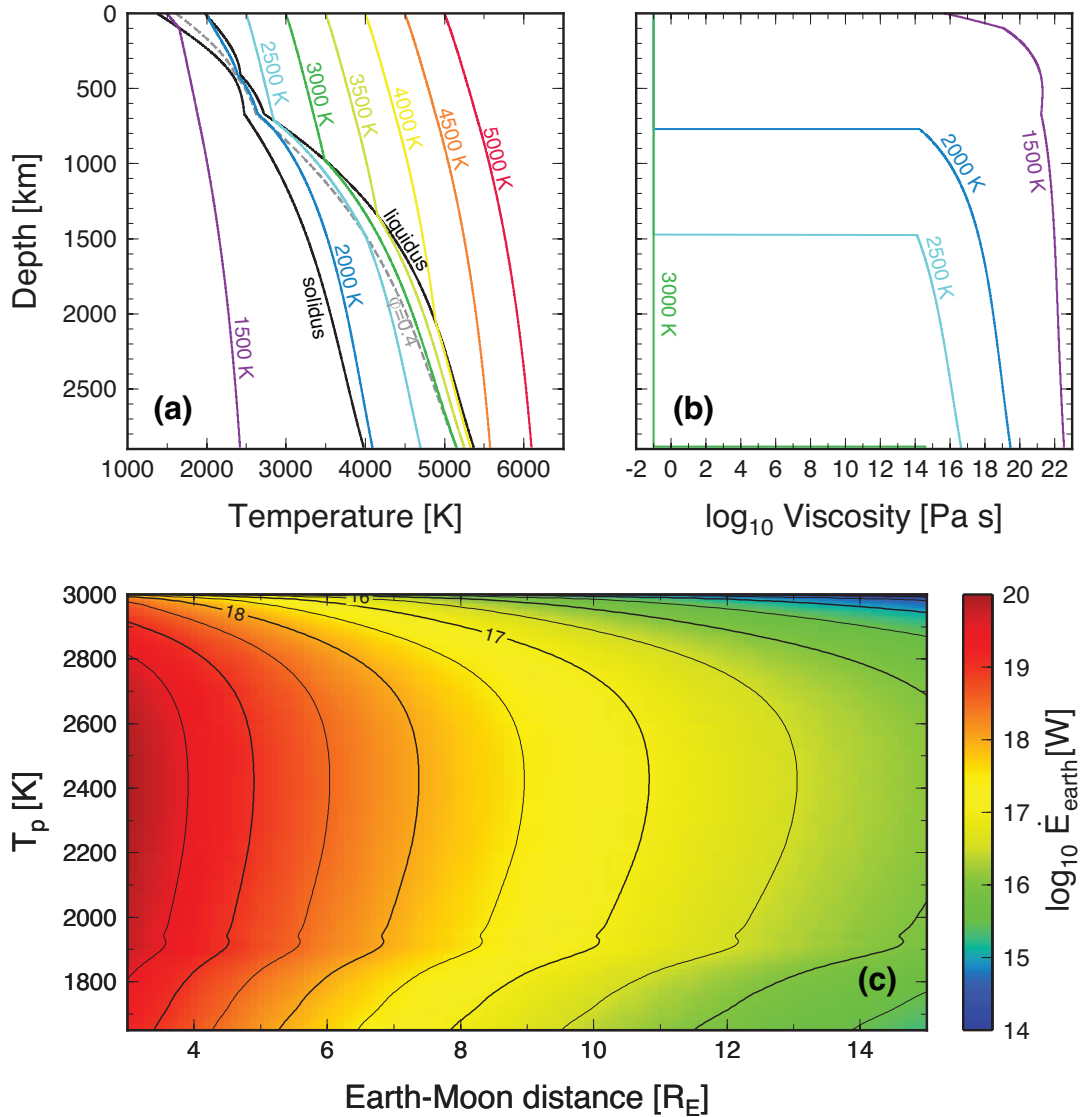,height=15cm}
\end{center}
\caption{\small (a) Mantle solidus and liquidus (solid), $T_{40}$
  (where the melt fraction is 0.4, the critical value for the
  rheological transition; gray dashed), and mantle adiabats with the
  potential temperatures $T_p$ from 1500~K to 5000~K, based on the
  `MK19' phase diagram. (b) Corresponding mantle viscosity structure
  with the reference background viscosity, $\alpha_{\eta}$=26, and the
  melt viscosity of 0.1~Pa~s; for $T_p>3000~K$, viscosity becomes that
  of melt at all depths as the adiabat is above $T_{40}$. (c)
  Dissipation of earth tides as a function of mantle potential
  temperature and the lunar distance. The tide-raising frequency
  varies with the distance, from $2.216 \times 10^{-4}$ rad~s$^{-1}$
  at 3~$R_E$ to $4.619\times 10^{-4}$ rad~s$^{-1}$ at 15~$R_E$.}
\label{fig:adiabat}
\end{figure}

\begin{figure}
\begin{center}
\epsfig{file=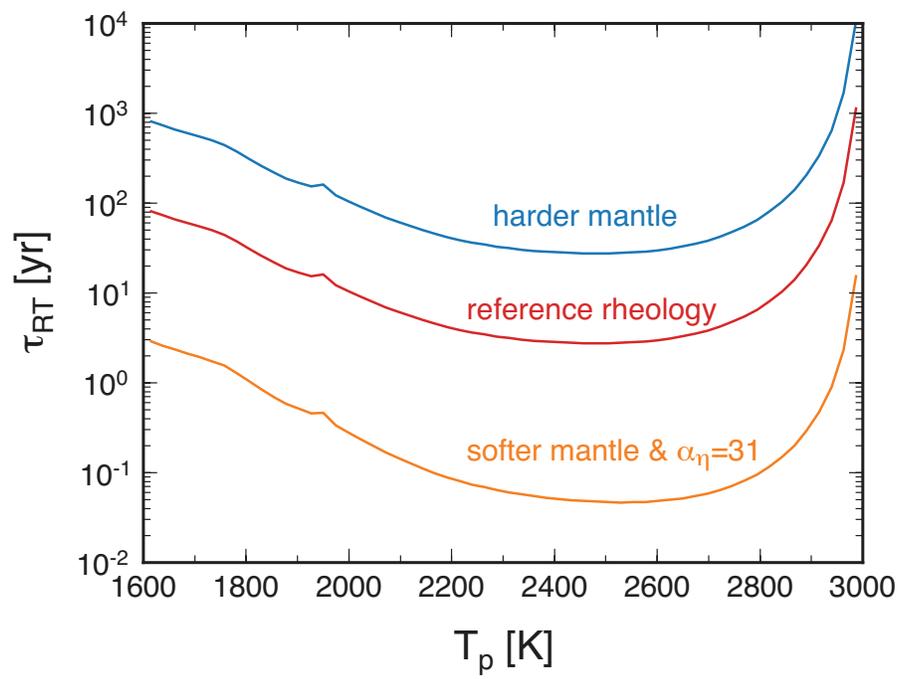,height=9cm}
\end{center}
\caption{\small Time scale for Rayleigh-Taylor instability as a
  function of mantle potential temperature, with three different
  settings for mantle rheology.}
\label{fig:RT}
\end{figure}

\begin{figure}
\begin{center}
\epsfig{file=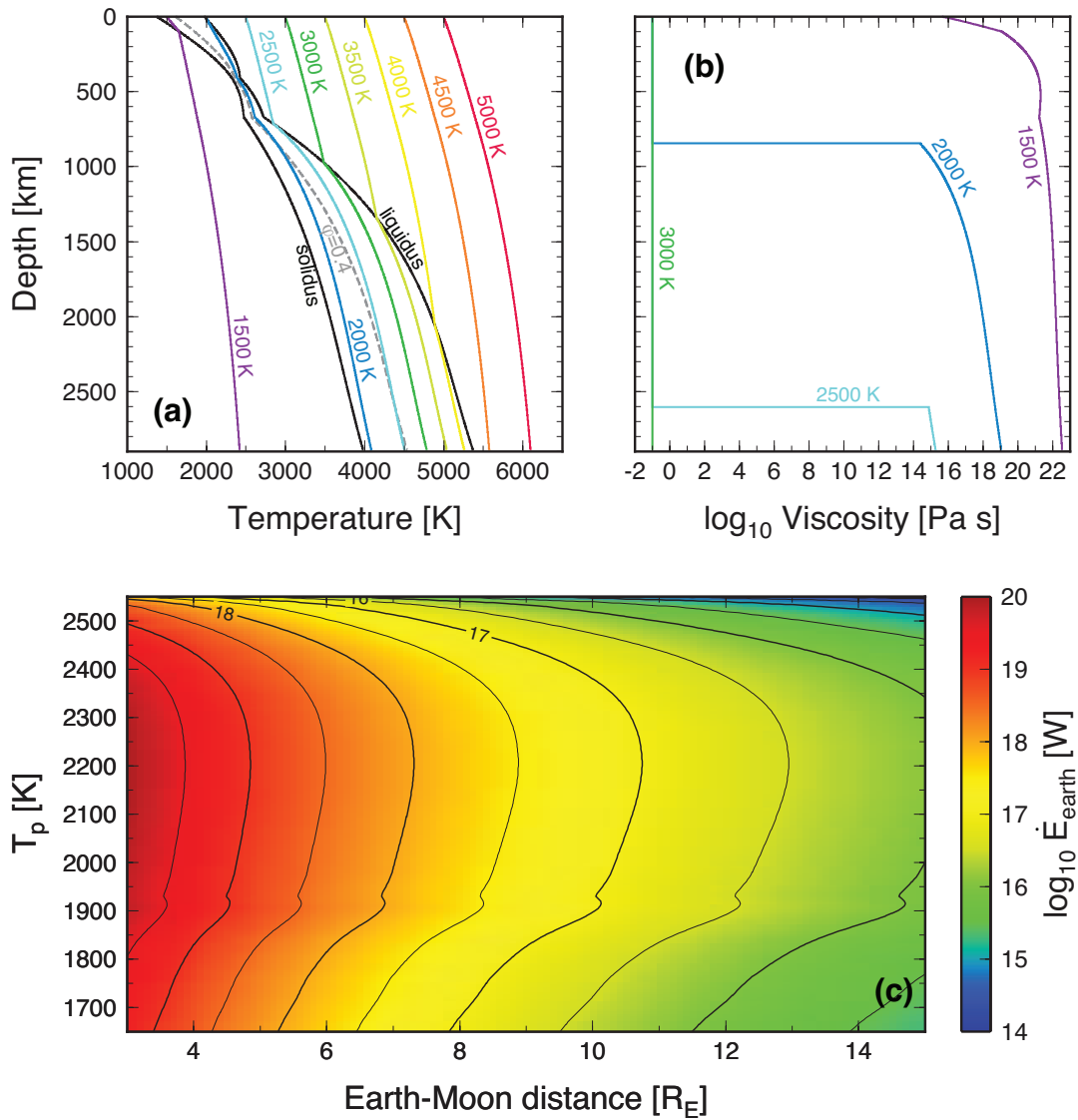,height=15cm}
\end{center}
\caption{\small Same as Figure~\ref{fig:adiabat}, but for the `const'
  phase diagram.}
\label{fig:adiabat2}
\end{figure}

\begin{figure}
\begin{center}
\epsfig{file=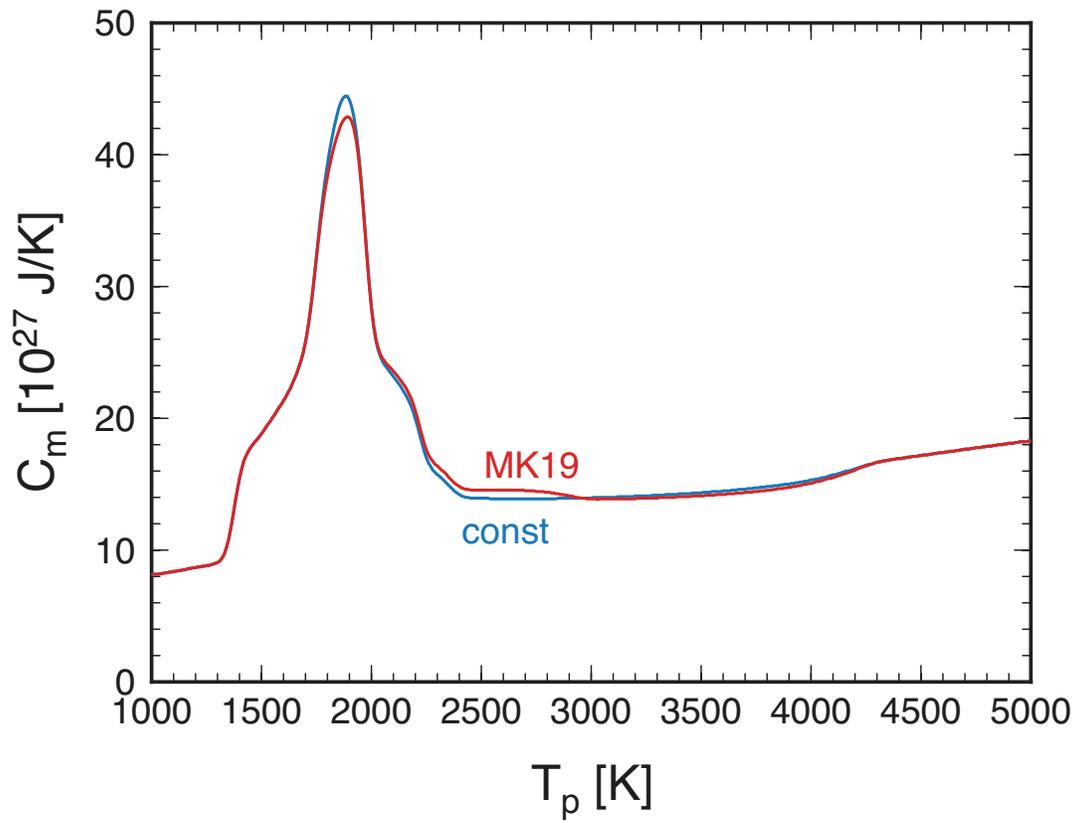,height=11cm}
\end{center}
\caption{\small Mantle heat capacity as a function of mantle potential
  temperature, with the `MK19' (red) and `const' (blue) phase
  diagrams.}
\label{fig:Cm}
\end{figure}

One of the unequivocal features of magma ocean solidification is that
it solidifies from the lower mantle (Figure~\ref{fig:adiabat}). This
is because the pressure dependence of the mantle liquidus is greater
than that of the mantle adiabat. Consequently, tidal dissipation in a
solidifying magma ocean takes place while the surface of the magma
ocean is well above the liquidus, releasing a large amount of heat. In
addition, the thermal structure of a solidifying magma ocean is always
close to adiabatic, as opposed to isothermal, because of the
Rayleigh-Taylor instability \cite[]{solomatov15tog, miyazaki19b}. This
limits the depth extent of optimal viscosity that is important for
tidal resonance (Figure~\ref{fig:adiabat}b), and thus the magnitude of
tidal dissipation (Figure~\ref{fig:adiabat}c).  As such, tidal
dissipation in a solidifying magma ocean is always substantially lower
than its surface heat loss, being ineffective to slow down the
solidification process. The completion of the magma ocean stage,
defined by the time when the surface melt fraction reaches to
$\sim$0.4, can take place within only $\sim$$10^5$ years
(Figure~\ref{fig:evolution}a), and the lunar distance is still at
$\sim$8~$R_E$ in the reference case (Figure~\ref{fig:evolution}c).

As noted earlier, carbon dioxide and water, two dominant volatile
species in Earth's magma ocean \cite[]{sossi20}, have contrasting
solubilities in magma, the latter being more soluble than the former
\cite[]{elkins-tanton08}. When its melt fraction is reduced to
$\sim$0.4, a solid-melt mixture experiences the rheological
transition, jumping from melt viscosity to solid-like viscosity
\cite[]{abe93b, solomatov15tog} (Figure~\ref{fig:adiabat}b). The
combination of the rheological transition and the Rayleigh-Taylor
instability through a solidifying magma ocean results in the efficient
entrapment of water in the mantle \cite[]{hier-majumder17,
  miyazaki22a}, and after the completion of the magma ocean stage, a
water ocean will form gradually first by the upward percolation of
remaining melt in the shallow upper mantle, and then by degassing
associated with subsolidus mantle convection \cite[]{miyazaki22a}.

As soon as a water ocean emerges, the effect of ocean tides needs be
considered, but given our limited understanding of the landscape of
the early Earth, it is still impractical to model the details of ocean
tides in the early Earth. To guide our discussion, therefore,
hypothetical orbital evolution with a range of tidal $Q/k_2$, where
$k_2$ is the Love number, is shown in Figure~\ref{fig:evolution}c. To
match the orbital reconstruction of \cite{tyler21} or \cite{daher21}
at the time of $10^7$ years, for example, $Q/k_2$ has to be
consistently as low as $\sim$2, which seems unrealistic. These studies
do not consider continental growth nor variable ocean volume in the
early Earth, so if we only aim to converge to their model predictions
at the end of the Hadean ($5 \times 10^8$ years) or the end of the
Archean ($2 \times 10^9$ years), $Q/k_2$ can take values that are
usually assumed in previous studies ($\sim$50-100)
\cite[]{ward00b,pahlevan15,cuk21}. It is important to note that the
details of lunar recession after the solidification of the magma ocean
($\sim$$10^5$ years) until the end of the Archean remain highly
uncertain, because the period corresponds to the part of Earth history
during which both continental mass and ocean volume can change
dramatically \cite[]{korenaga21c} and because tidal dissipation in the
ocean can be very sensitive to such changes
\cite[]{motoyama20}. Nevertheless, the Hadean tidal environment is
unlikely to have been intense enough to challenge the possibility of
wet-dry cycles on stable landmasses. The tidal amplitude is inversely
proportional to the cube of the lunar distance; with the present-day
value of $\sim$0.5~m, the amplitude can be up to only several meters
after $\sim$4.3~Ga (i.e., after 200 Myr in Figure~\ref{fig:LOD}).

It is difficult to increase the extent of lunar recession during the
magma ocean stage (Figures~\ref{fig:evolution}c and
\ref{fig:dist}). The maximum tidal $Q$ during this stage is set to
$10^4$ in the reference case (Methods), which is already at the lower
end of the acceptable range \cite[]{zahnle15}. Starting closer to
Earth, i.e., with greater tidal dissipation, has only a self-limiting
effect (Figure~\ref{fig:evolution}c). Varying mantle rheology within a
reasonable range of its uncertainty can amount to a difference of only
$\sim$$\pm$0.5~$R_E$ (Figure~\ref{fig:dist}). The lunar distance at
the end of the magma ocean stage can be pushed up to $\sim$8.5~$R_E$
with a softer mantle background, but this choice of background
viscosity may not be appropriate given the partitioning behavior of
water \cite[]{karato86b} (see the subsection ``Rheology''). The lunar
distance at the end of the magma ocean stage is most limited in the
case of the harder mantle with the initial distance of 3.5~$R_E$
(dashed blue curve; $\sim$7.5~$R_E$). The effect of phase diagram
choice, whether `MK19' or `const', is trivial. The most effective way
is to increase melt viscosity from a peridotitic value (0.1~Pa~s
\cite[]{dingwell04}) to a basaltic one (10~Pa~s \cite[]{mcbirney93}),
which lowers surface heat flux by a factor of $\sim$5 and thus
lengthens the duration of the magma ocean stage. Fractional
crystallization, if it takes place, does modify the composition of
surface magma, but the composition remains largely peridotitic during
the bulk of the solidification \cite[]{miyazaki19b}; thus, the case
with basaltic melt viscosity (Figures~\ref{fig:evolution}c and
\ref{fig:dist}) should be regarded as an unlikely end-member.

The effect of varying the critical melt fraction $\phi_c$ is also minor
on lunar recession (Figure~\ref{fig:phic}). Also, the use of Maxwell
viscoelasticity leads to more reduced lunar recession as expected
(Figure~\ref{fig:phic}). Our calculation of orbital evolution does
not take into account the effect of tidal dissipation within the Moon.
However, doing so would only result in reducing the lunar
semimajor axis \cite[]{murray00book}, thereby strengthening the case
of limited lunar recession during magma ocean solidification.

\section{Discussion}

\begin{figure}
\begin{center}
\epsfig{file=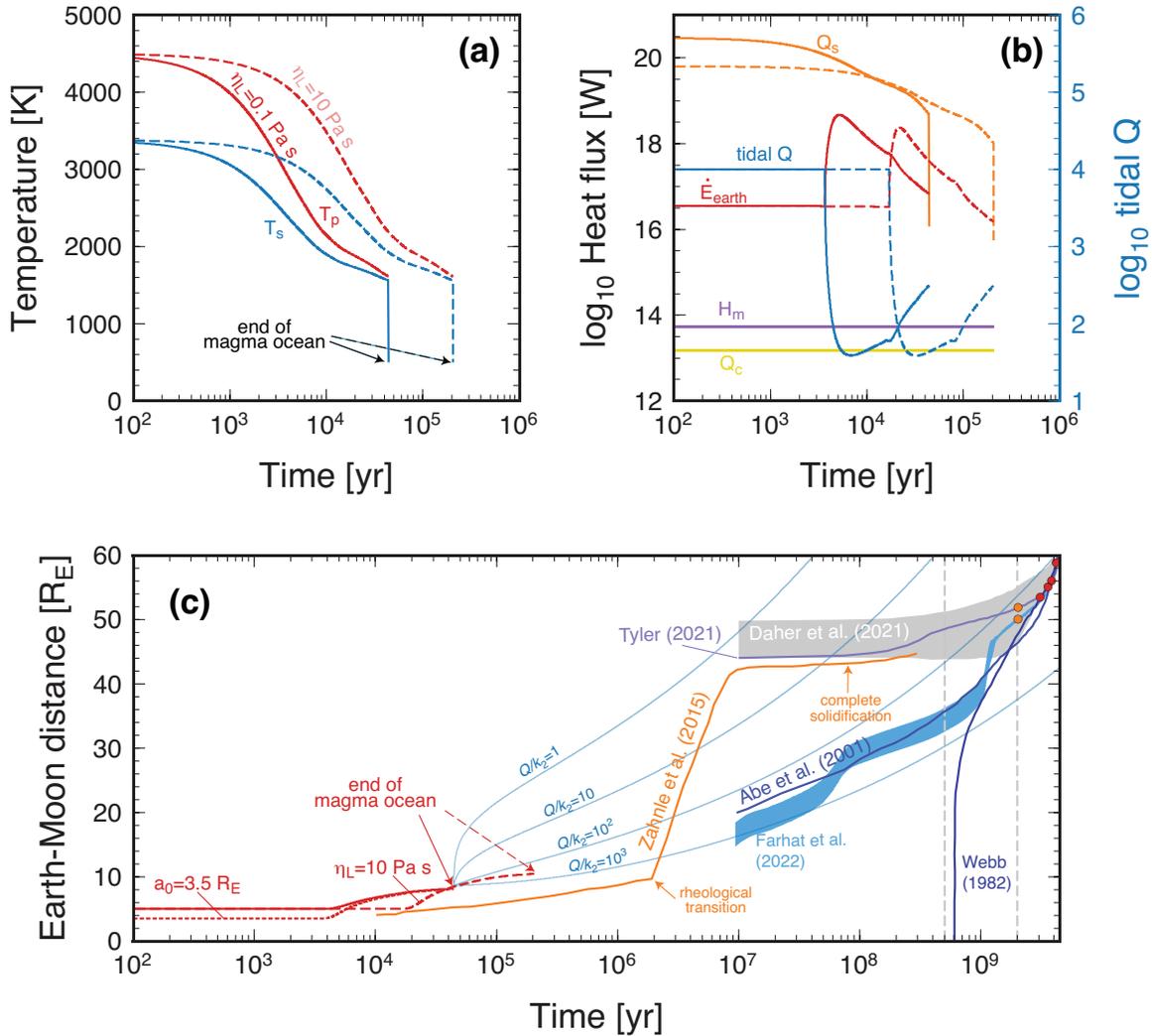,height=14cm}
\end{center}
\caption{\small (a) Evolution of mantle potential temperature $T_p$
  (red) and surface temperature $T_s$ (blue) for the reference case
  (solid), which uses the `MK19' phase diagram, standard background
  viscosity, $\alpha_{\eta}=26$, the melt viscosity of 0.1~Pa~s,
  maximum tidal $Q$ of $10^4$, and the initial lunar distance of
  5~$R_E$, starting at $T_p$ of 4500~K. The case with the same model
  setting except for the melt viscosity of 10~Pa~s is also shown
  (dashed). (b) Corresponding evolution of surface heat flux $Q_s$
  (orange), tidal dissipation $\dot{E}_{\mbox{\scriptsize earth}}$
  (red), tidal $Q$ (blue), radiogenic heat production $H_m$ (purple),
  and core heat flux $Q_c$ (yellow). The case with the melt viscosity
  of 10~Pa~s is in dashed. (c) Evolution of the lunar distance for the
  reference case (red), and those with the melt viscosity of 10~Pa~s
  (red dashed) and with the initial distance of 3.5~$R_E$ (red
  dotted). For the reference case, the evolution after the magma ocean
  phase (blue) is shown with a constant $Q/k_2$ (1 to $10^3$).
  Previous model predictions are also shown for comparison (orange:
  \cite{zahnle15}, dark blue: \cite{webb82}, blue: \cite{M_abe01},
  purple: \cite{tyler21}, gray: \cite{daher21}, and light blue:
  \cite{farhat22}). For the models of \cite{webb82}, \cite{M_abe01},
  \cite{tyler21}, and \cite{daher21}, the start time of the Earth-Moon
  system is set to 4.5~Ga, whereas it is assumed to be 4.429~Ga in the
  model of \cite{farhat22}. These model predictions are shown only
  after $10^7$~years as they are not expected to be accurate for the
  very early Earth. In the model of \cite{zahnle15}, the whole mantle
  experiences the rheological transition at $\sim$$2 \times 10^6$
  years and completely solidifies at $\sim$$8 \times 10^7$ years, as
  indicated by arrows. Small circles denote geological constraints
  from tidal (red) and banded iron (orange) deposits
  \cite[]{coughenour09, meyers18, lantink22}.}
\label{fig:evolution}
\end{figure}

\begin{figure}
  \vspace*{3cm}
\begin{center}
\epsfig{file=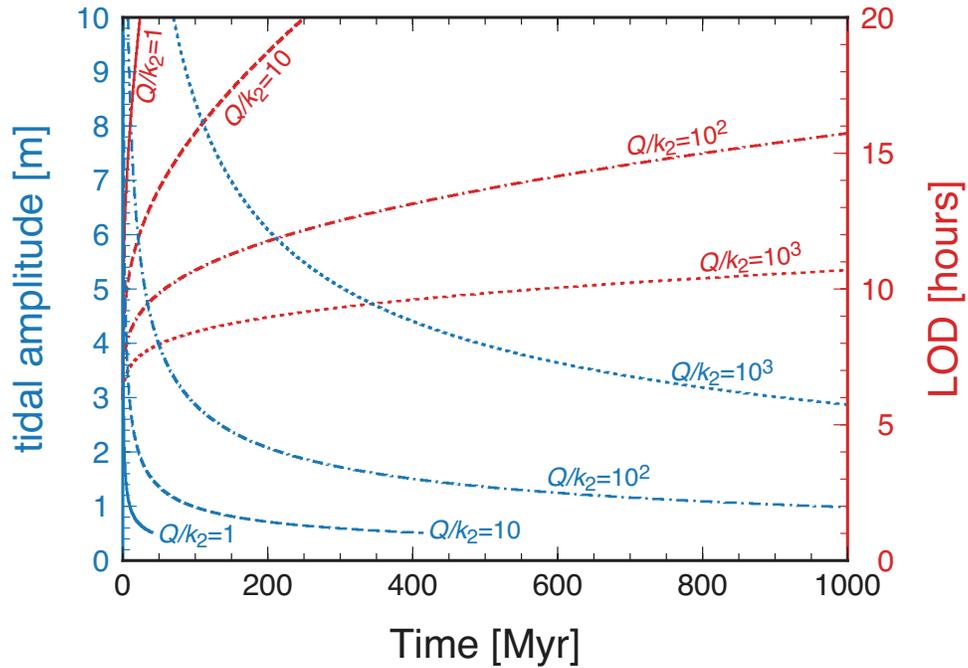,height=9cm}
\end{center}
\caption{\small Tidal amplitude, with respect to the reference
  amplitude of 0.5~m at present (blue), and length of the day (red)
  for Earth's first one billion years, with a range of tidal $Q/k_2$:
  1 (solid), 10 (dashed), $10^2$ (dot-dashed), and $10^3$ (dotted).}
\label{fig:LOD}
\end{figure}

\begin{figure}
\begin{center}
\epsfig{file=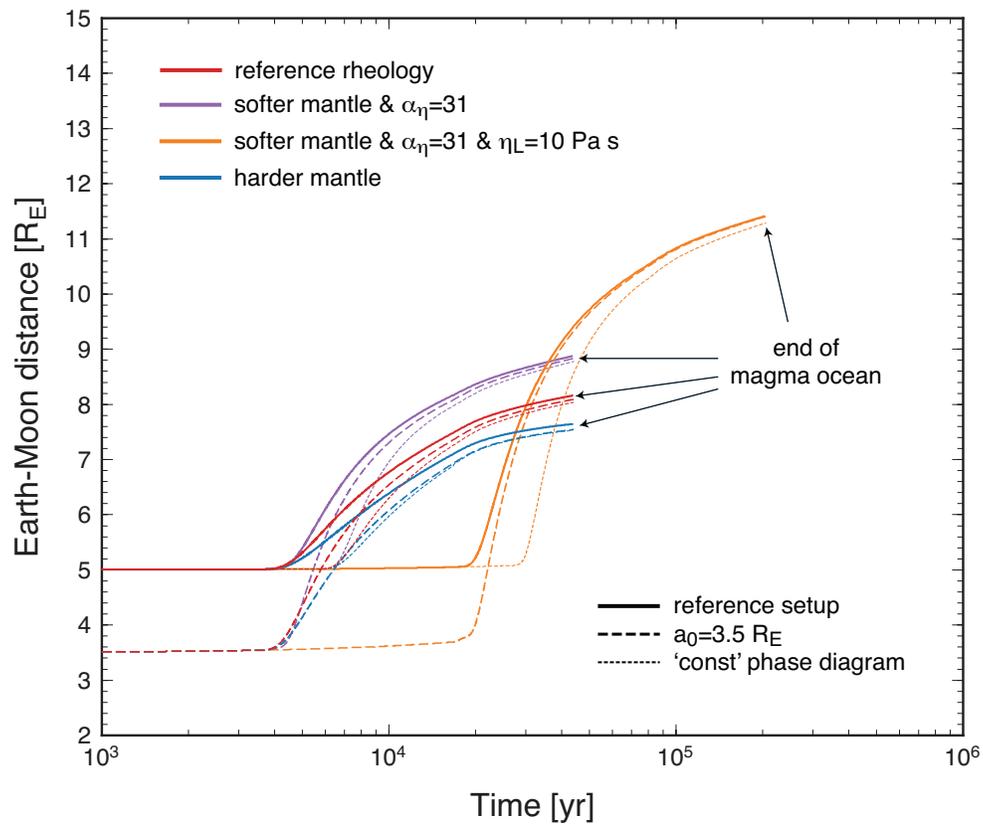,height=11cm}
\end{center}
\caption{\small Early history of lunar recession with a range of
  parameter combinations: four choices of mantle rheology (red -
  reference, purple - softer background with $\alpha_{\eta}=31$
  (i.e., softest solid viscosity), orange - softer background with
  $\alpha_{\eta}=31$ and melt viscosity of 10~Pa~s, and blue - harder
  background), the initial lunar distance of 5~$R_E$ (solid) and
  3.5~$R_E$ (dashed), and the `MK19' (solid) and `const' (dotted)
  phase diagrams.}
\label{fig:dist}
\end{figure}

\begin{figure}
\begin{center}
\epsfig{file=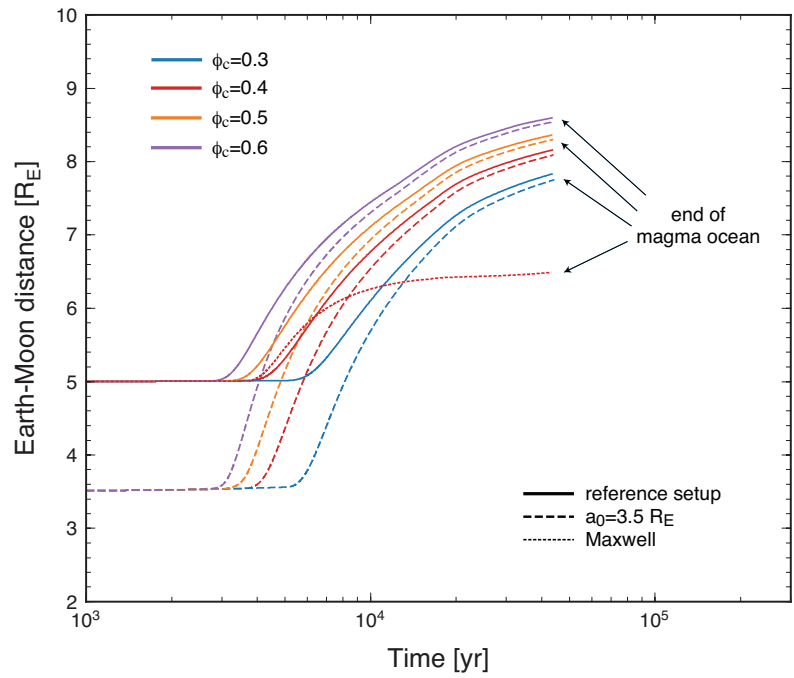,height=9cm}
\end{center}
\caption{\small Early history of lunar recession with a range of
  critical melt fraction $\phi_c$: 0.3 (blue), 0.4 (red), 0.5
  (orange), and 0.6 (purple) and with the initial lunar distance of
  5~$R_E$ (solid) and 3.5~$R_E$ (dashed). The other model parameters
  are the same as the reference case. The case of Maxwell body with
  $\phi_c=0.4$ is also shown in dotted for comparison.}
\label{fig:phic}
\end{figure}

\begin{figure}
\begin{center}
\epsfig{file=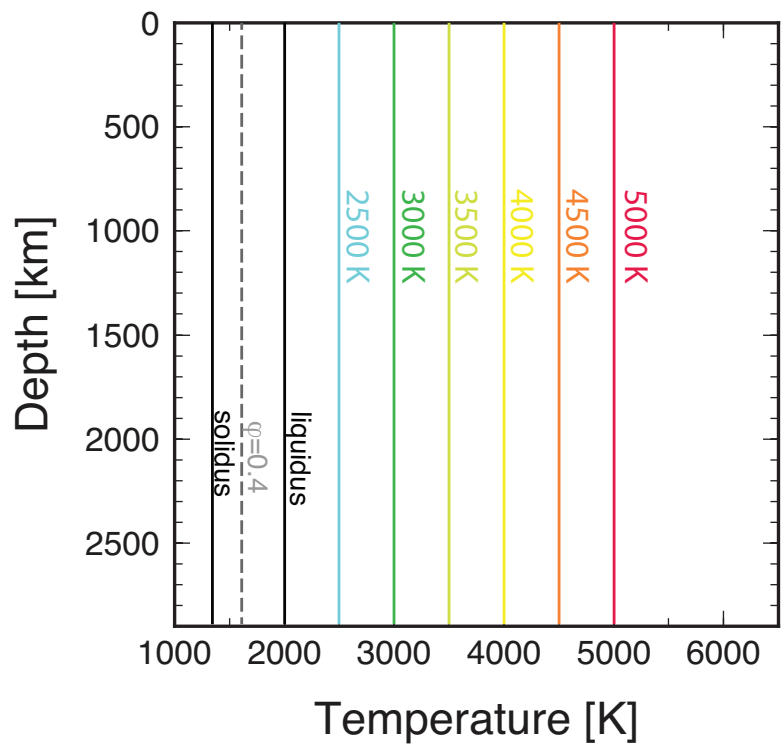,height=10cm}
\end{center}
\caption{\small The phase diagram of the mantle and `adiabats' assumed in
  \cite{zahnle15}.}
\label{fig:iso}
\end{figure}

Differences from the model of \cite{zahnle15}, which is also
based on the time-forward approach with magma ocean solidification,
are striking (Figure~\ref{fig:evolution}c), and most of them can be
traced back to their isothermal approximation. This approximation was
necessary for their analysis because their calculation of tidal
dissipation assumed homogeneous material properties for the entire
Earth. In our model, a magma ocean starts to experience the
rheological transition and a significant amount of tidal dissipation
as soon as the surface temperature drops below $\sim$3000~K
(Figure~\ref{fig:adiabat}), but with the isothermal approximation, in
which the solidus and liquidus temperatures do not change with depth
(Figure~\ref{fig:iso}), such a transition has to wait until the
surface melt fraction reaches $\sim$0.4, which happens at the surface
temperature of $\sim$1560~K in their model. Such low magma temperature
limits surface heat flow, and at the same time, tidal dissipation is
maximized as the entire mantle has low viscosities around the
rheological transition. Tidal heating is thus effectively trapped in
Earth's magma ocean, extending its lifetime by tens of million years,
during which substantial lunar recession can take place
(Figure~\ref{fig:evolution}c). The isothermal approximation introduces
about three orders of magnitude difference in the solidification time
scale, which results in a factor of $\sim$5 difference in the
Earth-Moon distance. This may not be surprising given the drastic
nature of the isothermal approximation (Figure~\ref{fig:iso}).

As discussed above, the isothermal approximation made by
\cite{zahnle15} maximizes the total amount of tidal dissipation (1) by
bringing the entire mantle to the rheological transition and (2) by
keeping the entire mantle close to the rheological transition for a
prolonged time with low surface heat flux. Whereas the isothermal
approximation (Figure~\ref{fig:iso}) is clearly inappropriate given
our understanding of the mantle phase diagram, and maintaining a
thermal profile at the rheological transition is dynamically
implausible because of the Rayleigh-Taylor instability, it is worth
exploring other possibilities to increase tidal dissipation within a
solidifying magma ocean and thus the extent of early lunar
recession. In our model, a magma ocean starts to solidify from the
bottom of the lower mantle (Figure~\ref{fig:adiabat}a), and because of
the Rayleigh-Taylor instability (Figure~\ref{fig:RT}), the thermal
structure of a solidifying magma ocean remains adiabatic. There are a
few possibilities to deviate from this scenario. First, even though
thermodynamic calculations suggest bottom-up solidification
\cite[]{miyazaki19b}, solidification from a mid-mantle level could
still occur using a thermodynamic database different from that of
\cite{miyazaki19a}, and if it happens, a basal magma ocean would form
\cite[]{labrosse07b, mosenfelder09, stixrude09}.  Even with bottom-up
solidification, a basal magma ocean could still form if matrix
compaction is efficient \cite[]{miyazaki19b}. A basal magma ocean
would solidify from the top, with a limited extent of a partially
molten boundary layer \cite[]{labrosse07b}, so tidal dissipation
associated with a basal magma ocean is unlikely to be significant.  In
general, matrix compaction (or equivalently, melt percolation)
increases the vertical extent of a fluid layer ($\phi>\phi_c$) and
reduces the melt fraction of other solid parts of the mantle ($\phi
\le \phi_c$). Tidal dissipation within a fluid layer is considerably
lower than that within a solid layer, even using the maximum fluid
tidal $Q$ of $10^4$ (Figures~\ref{fig:evolution}b), and reduction in
the melt fraction drives the solid layer away from the rheological
transition, which acts to reduce tidal dissipation. In other words,
the solidification scenario of our model probably corresponds to the
case of the maximum tidal dissipation, and the most important finding
of this study is that, even with this optimal scenario, lunar
recession during magma ocean solidification is severely limited.

Limited lunar recession during the early Hadean, seen in our model
results, may resurrect the early origin of the lunar inclination,
either by the evection resonance \cite[]{touma98} or by interaction with
remanent proto-lunar disk materials \cite[]{ward00b}. The present-day
lunar orbit has an inclination of $\sim$5$^{\circ}$ with respect to
the ecliptic plane, and the origin of this inclination has been
the most serious dynamical problem for the formation mechanism of the
Moon \cite[]{goldreich66, touma98, ward00b, pahlevan15, cuk16,
  ZL_tian20}. To preserve the primordial lunar inclination, it is
suggested that the lunar magma ocean has to solidify before the onset
of the Cassini state transition at $\sim$30~$R_E$ \cite[]{EMA_chen16}.
As the lunar magma ocean is generally considered to have solidified
with a time scale of 10-100~Myr \cite[]{elkins-tanton11, EMA_chen16},
such preservation appears impossible with the lunar recession model of
\cite{zahnle15} (Figure~\ref{fig:evolution}c). A proper treatment
of magma ocean solidification, however, suggests that lunar recession
is limited to only $\sim$8~$R_E$ at the end of the magma ocean phase,
and the preservation of primordial lunar inclination is certainly
within the bounds of possibility (Figure~\ref{fig:evolution}c).

The late origin of the lunar inclination, by collisionless encounters
with left-over planetesimals \cite[]{pahlevan15}, remains viable
although the efficiency of collisionless excitation is likely to be
reduced. This mechanism, which operates on the time scale of $10^8$
years, is more efficient for a larger cross-section of the Earth-Moon
system, and with the recession model of \cite{zahnle15}, the lunar
distance is already at $\sim$40~$R_E$ at the time of $10^7$ years.  In
light of our study, such recession requires an unrealistically low
$Q/k_2$ for an ocean-covered Earth. With more likely values of $Q/k_2$
($\sim 100$), the efficiency of collisionless excitation would be
reduced by $\sim$50\% than originally suggested
(Figure~\ref{fig:evolution}c).  Perhaps most important is that, with
limited early lunar recession, the early and late origins of the lunar
inclination are not mutually exclusive; all of them could contribute,
facilitating the resolution of the lunar inclination problem.

It is more difficult to assess the impact of our results on the lunar
orbital evolution with high initial obliquity \cite[]{cuk16, cuk21}. The
duration of such a scenario extends over $\sim$100~Myr, so most of the
suggested evolution must take place in the presence of a water ocean
(Figure~\ref{fig:evolution}c). It is common to use constant $Q/k_2$
for tidal dissipation in previous studies on lunar orbital evolution
\cite[]{ward00b,pahlevan15,cuk16,cuk21}, but ocean tides are expected to
have been highly variable on the early Earth \cite[]{motoyama20}, owing
to the competing effects of the growing ocean and continents
\cite[]{korenaga21c}. For the high initial obliquity scenario to work,
the Earth-Moon system needs to remove a considerable fraction of the
initial angular momentum through a series of resonances \cite[]{cuk21},
but the removal of angular momentum by resonance is efficient only for
a narrow range of tidal parameters \cite[]{rufu20}. A more incisive
evaluation of the lunar orbital evolution and the formation mechanism
of the Moon, therefore, hinges on an improved understanding of how
early surface environments were shaped by the dynamics of Earth's
interior.

\medskip
 
{\small

}

\section*{Acknowledgments}
The author thanks Brian Arbic and an anonymous reviewer, whose
comments helped to improve the accuracy of the manuscript. This
research is supported by the U.S. National Aeronautics and Space
Administration under Cooperative Agreement No.\ 80NSSC19M0069 issued
through the Science Mission Directorate.

\clearpage

\end{document}